\documentclass[10pt,aps,pra,twocolumn,preprintnumbers,amsmath,amssymb,superscriptaddress]{revtex4-2}
\UseRawInputEncoding
\usepackage{comment}
\usepackage{latexsym,amsthm,epsfig}
\usepackage[left=2.00cm, right=2.00cm, top=2.00cm, bottom=2.00cm]{geometry}
\usepackage[utf8]{inputenc}
\usepackage{comment}
\usepackage{mathrsfs}  
\usepackage{hyperref}
\hypersetup{
    citecolor=red,
    colorlinks=true,
    linkcolor=blue,
    filecolor=blue,
    urlcolor=blue,
}

\begin{document}
\title{Photonic Spin Hall Effect using bilayer Graphene in Nano Optomechanical Cavities}

\author{Muqaddar Abbas}
\altaffiliation{These authors contributed equally to this work.}
\affiliation{Ministry of Education Key Laboratory for Nonequilibrium Synthesis and Modulation of Condensed Matter,
Shaanxi Province Key Laboratory of Quantum Information and Quantum Optoelectronic Devices, School of Physics, Xi’an Jiaotong University, Xi’an $710049$, China}

\author{Muhammad Awais Altaf}
\altaffiliation{These authors contributed equally to this work.}
\affiliation{Department of Physics, University of Mianwali, Mianwali $42200$, Pakistan}

\author{Pei Zhang}
\email{zhangpei@mail.ustc.edu.cn}
\affiliation{Ministry of Education Key Laboratory for Nonequilibrium Synthesis and Modulation of Condensed Matter,
Shaanxi Province Key Laboratory of Quantum Information and Quantum Optoelectronic Devices, School of Physics, Xi’an Jiaotong University, Xi’an $710049$, China}
\author{Muhammad Waseem}
\email{mwaseem328@gmail.com}
\affiliation{Deparment of Physics and Applied Mathematics, Pakistan Institute of Engineering and Applied Sciences (PIEAS), Nilore $45650$, Islamabad, Pakistan}
\affiliation{Center for Mathematical Sciences, PIEAS, Nilore, Islamabad $45650$, Pakistan}

\date{\today}

\begin{abstract}
We propose a theoretical model to obtain the photonic spin Hall effect (SHE) in an optomechanical nanocavity using a graphene bilayer as the intracavity medium. In our model, the pump and probe fields coherently drive the first mirror, whereas the second mirror has mechanical oscillation due to the radiation pressure. We show that the right- and left-circular polarization components of the Gaussian probe field striking at an arbitrary incident angle become spatially separate along a direction orthogonal to the plane of incidence. Photonic SHE can be coherently controlled by adjusting the optomechanical interaction, cavity field and G-mode phonon coupling, as well as G-mode phonon and electronic state interaction. The findings of photonic SHE are equally valid for standard optomechanical systems in the absence of cavity field and G-mode phonon coupling and electronic state interaction. The cavity field and G-mode phonon coupling broadened the detuning range of the probe field to observe the dominant photonic SHE. Adding G-mode phonon and electronic state interaction generates enhanced photonic SHE at three different probe field detunings due to optomechanical-induced transparency being split into three windows. We show that asymmetric photonic SHE can be controlled through cavity field and G-mode phonon coupling and G-mode phonon and electronic state interaction when probe field detuning is non-zero. The photonic SHE in bilayer graphene integrated with an optomechanical cavity may enable further studies of spin-dependent photonic effects and quantum sensing applications.

\end{abstract}
\maketitle
\newpage
\section{Introduction}
The photonic spin Hall effect (SHE) is a fundamental phenomenon in light-matter interactions that creates the transverse spatial separation of light at the interface based on its polarization states \cite{onoda2004hall}. This effect emerges due to the spin-orbit interaction of light \cite{dBliokh2015}, where the polarization of photons plays an analogous role to the electron spin in conventional electronic spin Hall effects \cite{onoda2004hall, bliokh2006conservation}. It mirrors the SHE observed in electronic systems, where the gradient in the refractive index plays a similar role to the electrical potential difference, but here, it substitutes the spin of electrons with the polarization of photons \cite{sinova2015spin, valenzuela2006direct}. The theoretical foundation of photonic SHE was first introduced by Onoda \textit{et al.} in 2004 \cite{onoda2004hall} and was later expanded by in-depth theoretical analysis by Bliokh \cite{bliokh2006conservation}. Later, in 2008, Hosten and Kwiat experimentally validated this effect using weak measurement techniques \cite{hosten2008observation}.

The photonic SHE is now widely recognized as stemming from the spin-orbit interaction of photons, in line with the fundamental principle of angular momentum conservation in light \cite{cardano2015spin, dBliokh2015, tunnelspin}. Several mathematical and practical techniques have been developed to enhance this effect, including weak value amplification, which significantly magnifies the transverse spin-dependent displacement associated with the photonic SHE \cite{cai2017quantized, chen2015modified}. 
This phenomenon has applications in high-precision optical metrology, quantum information processing, and the development of advanced photonic devices for optical sensing and imaging \cite{aiello2009transverse, yin2013photonic, dong2024enhanced}. 
The photonic SHE has enabled applications such as probing topological phase transitions~\cite{shah2021probing, kort2017topological}, identifying graphene layers~\cite{zhou2012identifying}, chiral molecular detection~\cite{tang2023optimal}, and performing mathematical operations and edge detection~\cite{zhu2019generalized}.
Recently, the photonic SHE gained significant attention for its ability to control spin-dependent behaviors of photons in various optical media, including one-dimensional photonic crystals composed entirely of all-dielectric metamaterials~\cite{WU2023415348}, grating waveguide structures~\cite{PhysRevB.107.165428}, multilayered metallo-dielectric heterostructures realized through the hybridization of Tamm and surface plasmon polaritons~\cite{Srivastava}, and, in particular, graphene layers~\cite{zhou2012identifying,cai2017quantized,SHAH2024107676}.

Bilayer graphene is composed of two layers of monolayer graphene in AB stacking order. Bilayer graphene exhibits remarkable electronic properties due to interlayer coupling and its tunable band structure under external electric or magnetic fields. 
Unlike monolayer graphene, which features a linear band dispersion at the Dirac point, bilayer graphene possesses a parabolic dispersion relation that can be further modulated to induce a tunable bandgap through an applied perpendicular electric field \cite{castro2009electronic, ohta2006controlling, mccann2006asymmetry}. This property enables precise control over charge carrier dynamics, making it suitable for optoelectronic devices \cite{zhang_direct_2009, lui2011observation,novoselov2005two, novoselov2006unconventional}, the valley Hall effect \cite{xiao2007valley, mak2014valley}, and spintronics applications \cite{zhang_direct_2009, taychatanapat2010electronic}.
Graphene's unique electronic and optical properties have led to extensive research on its interaction with light, including the realization of the optical spin Hall effect \cite{nalitov2015spin}, Fano resonances~\cite{tang2010tunable}, ultrafast graphene-based optical switching \cite{ni2016ultrafast}, and the strong coupling of surface plasmon polaritons in monolayer graphene arrays \cite{wang2012strong}. 
Recently, the behavior of the photonic SHE was explored in bilayer graphene~\cite{Qi_2022}, bilayer graphene moiré superlattices~\cite{kort_PhysRevB}, and twisted bilayer graphene~\cite{Chen:21}.

However, coherent quantum properties of cavity optomechanical systems have drawn considerable attention in the last decades due to their application in quantum information, quantum sensing, optomechanical devices, and fundamental physics~\cite{RevModPhys.86.1391, chiu2008atomic, chaste2012nanomechanical,stapfner2013cavity,agarwal2010electromagnetically}. 
Furthermore, some recent studies integrated graphene sheets with nanomechanical resonators~\cite{chen2009performance, song2012stamp, bunch2007electromechanical, song2014graphene, weber2014coupling, hafeez2019optomechanically}. 
These studies revealed some interesting findings, such as Fano resonances~\cite{hafeez2019optomechanically}, thermal motion and back action cooling~\cite{song2014graphene}, tunable phonon-cavity coupling~\cite{de2016tunable}, force sensitivity~\cite{weber2016force}, tunable parametric amplification~\cite{Su_2021}, and entanglement~\cite{Ribeiro_2017}.

Very recently, graphene-based optomechanically induced transparency (OMIT) for resonant excitation of a surface plasmon polariton has been explored using only monolayer graphene or suspended membranes~\cite{akhter}.
OMIT is a fundamental mechanism underlying the realization of the photonic SHE \cite{tunnelspin, waseem2024gain, khan_loss-free_2025}. However, prior investigations, including Ref.\cite{akhter}, have not explored the photonic SHE in graphene-based cavity optomechanical systems, which is the main focus of the present paper.
The findings in Ref.~\cite{akhter} are limited to the observation of single-window OMIT in monolayer graphene-based nanocavities. In this study, we show the emergence of multiple OMIT windows, specifically, three transparency windows in bilayer graphene arising from phonon–exciton interactions.
As a consequence of this coupling, we find the appearance of a three-band photonic SHE, along with a pronounced asymmetry in the photonic SHE at finite probe detunings.
Moreover, we show that photonic SHE can be coherently controlled by tuning the optomechanical interaction ($g_{mc}$), cavity field and G-mode phonon coupling ($g_{cp}$), and G-mode phonon and electronic state interaction ($\lambda_k$). 
In the absence of $g_{cp}$ and $\lambda_k$, our results reduces to the case of standard optomechanical configuration~\cite{agarwal2010electromagnetically}, highlighting the broader relevance of our findings.

The structure of the paper is organized as follows. 
Section II provides a discussion on the physical system and its theoretical framework.
In Sec. III, we present and analyze the numerical results of our study. 
Finally, Sec. IV concludes the paper by summarizing the key results.
\begin{figure}
\begin{tabular}{@{}cccc@{}}
\includegraphics[width=1.0\linewidth]{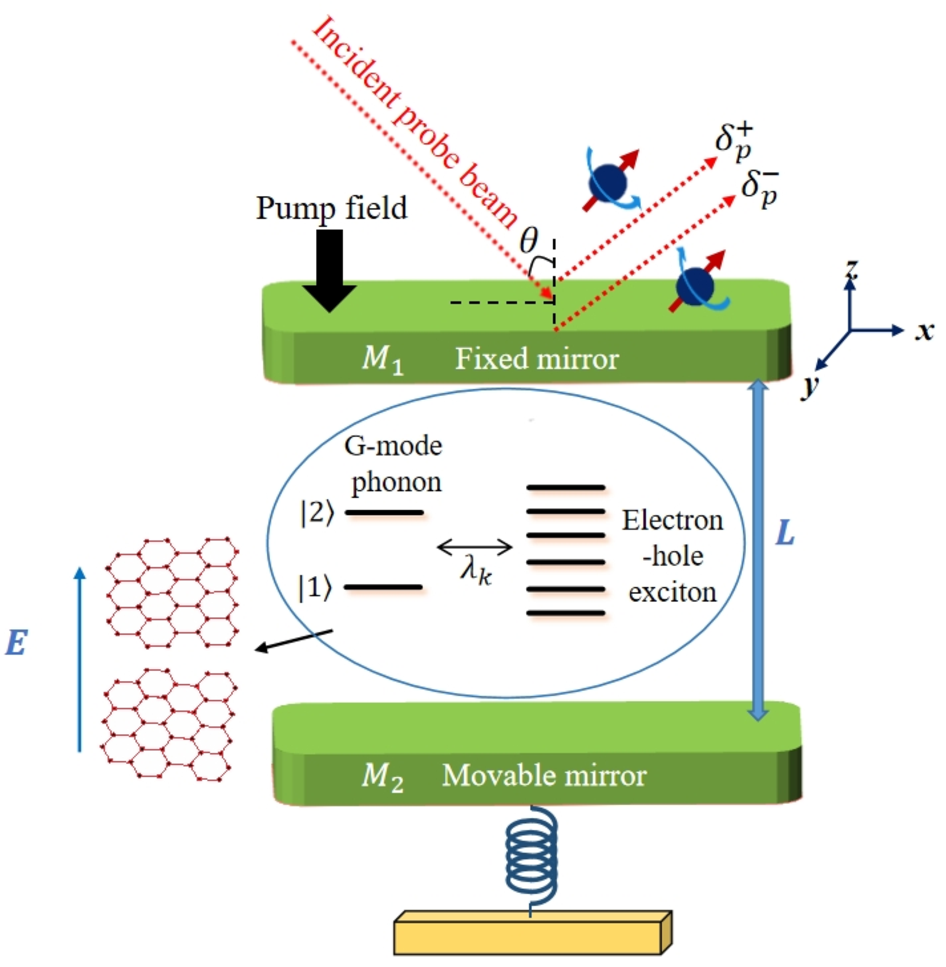}
\end{tabular}
\caption{Schematic diagram of a dual-gated bilayer graphene~\cite{tang2010tunable} integrated into an optomechanical cavity system, where $M_1$ represents a partially transmitted fixed mirror and $M_2$ is a fully reflective movable mirror.
A probe light beam strikes from a vacuum onto the cavity mirror $M_{1}$ at an angle $\theta$. 
Upon reflection, the left- and right-circular polarization components spatially separate along an axis orthogonal to the plane of incidence (the $y$ axis).
}
\label{figure1}
\end{figure}
\section{Physical System}
The physical setup is depicted in Fig. \ref{figure1}, which contains a bilayer graphene sheet inside an optomechanical nano cavity of length $L$. The mechanical component of the optomechanical system is represented by an ultrathin, perfectly reflecting mirror $M_2$, modeled as a quantum harmonic oscillator with a vibrational frequency $\omega_{b}$. The optical cavity supports a single isolated resonant mode with frequency $\omega_{c}$. In addition, the cavity is driven by the pump field of frequency $\omega_l$. The cavity field interacts with discrete two-level G-mode phonons, and phonons are coupled to electron-hole excitation through coupling $\lambda_k$. In our case, we have three layers: first is the partially reflecting mirror $M_1$ with wall thickness $d_1$ and permittivity $\epsilon_{1}$; second is the intracavity medium that consists of two-level state corresponding to the G-mode phonon, coupled to an electron cloud, with continuous modulation of electron-hole excitations through electrical gating with permittivity $\epsilon_{2}$; and third layer is mirror $M_2$ with thickness $d_2$ and permittivity $\epsilon_{3}$.

A probe light beam, consisting of both transverse electric (TE)- and transverse magnetic (TM)-polarized components, strikes from a vacuum onto the cavity mirror $M_{1}$ at an angle $\theta$, as shown in Fig. \ref{figure1}. 
Upon reflection at the system interface, the left- and right-circular polarization components spatially separate along an axis orthogonal to the plane of incidence (the $y$ axis), as illustrated in Fig. \ref{figure1}. 
Additionally, the applied electric field $E$ within the dual-gated bilayer graphene optomechanical cavity breaks symmetry, modifying both electronic and vibrational properties. This results in the opening of a bandgap and alters phonon dipoles, thereby influencing the overall optical and mechanical interactions within the system.\\

The complex reflection coefficients for TM polarization ($r_{p}$) and TE polarization ($r_{s}$) for the three-layer structure can be computed using the transfer matrix approach. The transfer matrix for the $i\text{th}$ layer of a given configuration for TE polarization \cite{Born1999, wu2010}
\begin{equation}
\mathcal{M}_{is}(k_x,\omega_p,d_i)=
\begin{pmatrix}
\cos[k_{z}^{i}d_i] & \dfrac{i\sin[k_{z}^{i}d_i]}{q_{is}} \\
iq_{is}\sin[k_{z}^{i}d_i] & \cos[k_{z}^{i}d_i]
\end{pmatrix}, \label{tra-mat}
\end{equation}
where $k_{z}^{i} = \sqrt{\epsilon_{i}k^2 - k^2\sin^2[\theta]}$ is the $z$ component of the wave vector in the $i\text{th}$ layer of the medium, and $d_i$ is the corresponding thickness of the $i\text{th}$ layer. The quantity $q_{is} = \sqrt{\epsilon_i k^2 - k^2\sin^2[\theta]}$ is used for the TE polarization case. In this context, $\epsilon_{i}$ is the permittivity of the $i\text{th}$ layer. The subscript $s$ indicates the TE polarization case. Furthermore, $k = 2\pi/\lambda$ is the wave vector in vacuum.

The total transfer matrix for the incident and reflected probe light beam for our proposed model can be written as
\begin{equation}\label{totaltransfermatrix}
\begin{split}
M^{\text{Total}}(k_x,\omega_p) =\; &\mathcal{M}_{1s}(k_x,\omega_p,d_1)\, \mathcal{M}_{2s}(k_x,\omega_p,d_2) \\
&\times \mathcal{M}_{3s}(k_x,\omega_p,d_3).
\end{split}
\end{equation}

Based on the total transfer matrix formalism, the TE-polarized reflection coefficient of the probe field is given by
\begin{equation}
r_s = \frac{q_{0s}(M^{\text{Total}}_{22} - M^{\text{Total}}_{11}) - (q_{0s}^2 M^{\text{Total}}_{12} - M^{\text{Total}}_{21})}{q_{0s}(M^{\text{Total}}_{22} + M^{\text{Total}}_{11}) - (q_{0s}^2 M^{\text{Total}}_{12} + M^{\text{Total}}_{21})},
\label{reflection_coefficient1}
\end{equation}
where $M^{\text{Total}}_{ij}$ are the elements of the total transfer matrix $M^{\text{Total}}(k_x, \omega_p)$, and the transverse wave vector component in the incident (and exit) medium (air) is defined as \( q_{0s} = \sqrt{\epsilon_0 k^2 - k^2 \sin^2 [\theta]} \).

Similarly, the TM-polarized reflection coefficient of the probe field is obtained by replacing \( q_{is} \) with \( p_{ip} \), and is given by
\begin{equation}\label{reflection_coefficient2}
r_p = \frac{p_{0p}(M^{\text{Total}}_{22} - M^{\text{Total}}_{11}) - (p_{0p}^2 M^{\text{Total}}_{12} - M^{\text{Total}}_{21})}{p_{0p}(M^{\text{Total}}_{22} + M^{\text{Total}}_{11}) - (p_{0p}^2 M^{\text{Total}}_{12} + M^{\text{Total}}_{21})},
\end{equation}
where \( p_{0p} = \frac{q_{0s}}{\epsilon_0} \) corresponds to the TM counterpart of the transverse wave vector in the air region~\cite{asiri_controlling_2016}.

For a Gaussian beam reflected by the interface, the field amplitudes of the two circular components of reflected light are arranged as
\begin{eqnarray}\label{field-amplitude}
    \mathcal{E}^{\pm}_{r}(x_{r},y_{r},z_{r})&=&\frac{w_0}{w}\text{exp}[-\frac{x^{2}_{r}+y^{2}_{r}}{w}]\times\nonumber\\&&[r_p-\frac{2ix_{r}}{kw}\frac{\partial r_p}{\partial\theta}\mp\frac{2y_{r}\text{cot}[\theta]}{kw}\nonumber\\&&(r_s+r_p)],
\end{eqnarray}
where $w=w_0[1+(2z_{r}/k_{1}w_0^{2})^{2}]^{1/2}$, $z_{r}=k_{1}w^2_{0}/2$ is the Rayleigh length, \(w_{0}\) represents the radius of the waist of the incident beam, $(x_{r},y_{r},z_{r})$ shown the coordinate system for reflected light, and the superscript $\pm$ represents the different spin states. Here, $k_{1}=\sqrt{\epsilon_{1}}k$. Then, the transverse displacement of reflected light can be expressed as
\begin{equation}\label{transverse-displacement}
\delta^{\pm}_{p}=\frac{\int y|\mathcal{E}^{\pm}_r(x_{r},y_{r},z_{r})|^2dx_{r}dy_{r}}{\int |\mathcal{E}^{\pm}_r(x_{r},y_{r},z_{r})|^2dx_{r}dy_{r}}.
\end{equation} 
From Eqs.(\ref{field-amplitude}) and (\ref{transverse-displacement}), the corresponding transverse spin-displacements components $\delta^{+}_{p}$ and $\delta^{-}_{p}$ in terms of refractive coefficients can be expressed as \cite{ xiang2017enhanced,zubari}

\begin{equation}
\delta^{\pm}_{p} = \mp \frac{k_{1} w_{0}^{2} \operatorname{Re}\left[1 + \frac{r_{s}}{r_{p}}\right] \cot [\theta]}
{ k_{1}^{2} w_{0}^{2} + \left| \frac{\partial \ln r_{p}}{\partial \theta} \right|^{2} + \left| \left(1 + \frac{r_{s}}{r_{p}}\right) \cot [\theta] \right|^{2} }.
\label{halleffect}
\end{equation}

Here, $\delta^{\pm}_{p}$ denotes the transverse shift between the left and right circularly polarized components of the incident light. In the following, we will concentrate on the transverse shift $\delta^{+}_{p}$ of the left circularly polarized component. Since the two spin components have equal magnitudes but opposite directions, the shift of the right circularly polarized component can be adjusted concurrently.

In our model, the permittivities of the cavity walls ($\epsilon_1$ and $\epsilon_3$) are held constant, while the permittivity of the intracavity medium $\epsilon_2$ is dynamically linked to the susceptibility $\chi$ via the relation $\epsilon_2 = 1 + \chi$. In atomic systems, the susceptibility $\chi$ characterizes the optical response of the probe field~\cite{scully1997quantum}.
In optomechanical systems, the output field $E_{\text{out}}$ plays an analogous role to $\chi$~\cite{agarwal2010electromagnetically}. Therefore, we define $\chi=E_{\text{out}}$~\cite{khan2020investigation,liu2024generating,waseem_ghsmagno}.

Reflection coefficients $r_s$, $r_p$, and photonic SHE depend on the tunability of $\epsilon_{2}$, and hence $\chi$. Therefore, next we calculate $\chi$ using the quantum Langevin approach. 
Therefore, we start by writing the total Hamiltonian of the system in the rotating frame of the pumping field $H=H_0+H_{\text{int}}+H_{dr}$.
The free part of the Hamiltonian is 
\begin{eqnarray}
H_0&=&\hbar\Delta_{c}c^{\dagger}c+\hbar\omega_{b}b^{\dagger}b+\hbar\Delta_{n}\sigma_{z}+\hbar\Delta_{ex}d_{k}^{\dagger}d_{k},\nonumber\\
\end{eqnarray}
where the first term on the right side with $c^{\dagger}$ and $c$ shows the creation along with annihilation operator of the cavity field, with $\Delta_c=\omega_c-\omega_l$ being the detuning of the cavity field frequency from the pump field frequency. 
The second term shows the creation (annihilation) operator of optomechanical vibrational modes $b^{\dagger}$ ($b$) with vibration frequency $\omega_{b}$. 
The Hamiltonian of a two-level system with a transition frequency of $\omega_n$ is represented by the third element. 
Here, $\Delta_{n}=\omega_n-\omega_p$ is the detuning of a discrete two-level G-mode phonon from probe field frequency $\omega_p$, whereas $\sigma_z$ is the Pauli spin operator for the two-level G-mode phonon. 
The last term denotes the electron-hole excitation energy where annihilation along with creation functions for the electron–hole excitement are denoted as $d_{k}$ and $d^{\dagger}_{k}$, respectively, and $\Delta_{ex}=\omega_{ex}-\omega_p$ is detuning between the electron-hole excitation frequency $\omega_{ex}$ and probe field frequency.

The interaction part of the Hamiltonian is 
\begin{eqnarray}
H_{\text{int}}&=&-\hbar g_{mc} c^{\dagger} c (b+b^{\dagger}) + \hbar g_{cp}(c^{\dagger}\sigma_{-}+c\sigma_{+}) \nonumber\\ &&-\hbar\lambda_{k}(\sigma_{-}d_{k}^{\dagger}+\sigma_{+}d_{k})
\end{eqnarray}
The first term pertains to the interaction between cavity modes and the oscillating mirror $M_2$, characterized by the coupling strength $g_{mc}=(\omega_c/L)\sqrt{\frac{\hbar}{2 m \omega_b}}$ \cite{agarwal2010electromagnetically}.
This coupling term quantifies how the motion of the mechanical oscillator influences the cavity mode, leading to optomechanical interactions.
The second term $g_{cp}$ is the linear coupling between the cavity field and G-mode phonon interaction, which means phonons are considered infrared active~\cite{tang2010tunable}.
Here, $\sigma_+$ and $\sigma_-$ function as raising and lowering operators for the two-level graphene G-mode phonon, respectively.
Similarly, the electron-phonon coupling strength, $\lambda_{k}$, represents the interaction between the G-mode phonon and the electronic states (electron-hole excitation) in bilayer graphene. This coupling is often described using the Huang-Rhys parameter, which characterizes the strength of phonon-assisted electronic transitions \cite{wei2023polaritonic}.

The Hamiltonian part $H_{dr}$ comprises classical fields, namely the pump and probe, which possess frequencies $\omega_l$ and $\omega_p$, respectively. It can be written as
\begin{eqnarray}
H_{dr}&=&i\hbar \mathcal{E}_l(c^{\dagger}-c)+i\hbar \mathcal{E}_p(c^{\dagger}e^{-i\delta t}-c e^{i\delta t}).\label{hamiltonian2}
\end{eqnarray} 
The amplitudes of the pump as well as probe light fields are specified as $\mathcal{E}_l=\sqrt{2\kappa P_l/\hbar\omega_l}$ and $\mathcal{E}_p=\sqrt{2\kappa P_p/\hbar\omega_p}$, respectively. 
Here, $P_l$ is the power of the pump field, $P_p$ is the power of the probe field, $\kappa$ is the cavity decay rate, and $\delta=\omega_p-\omega_l$.

Next, we employ the Heisenberg equation of motion to solve the dynamics of the system's observable, which results in the following set of equations 
\begin{eqnarray}
\dot c&=&[-i\Delta_c+i g_{mc} (b+b^{\dagger}) -\kappa]c-i g_{cp}\sigma_{-}+\mathcal{E}_l+\mathcal{E}_pe^{-i\delta t}\nonumber\\
&+&\sqrt{2\kappa}c_{in}(t), \nonumber\\
\dot b &=& (-i\Delta_b - \Gamma_m) b + ig_{mc} c^{\dagger}c + \zeta(t),\nonumber\\
\dot \sigma_{-}&=&-(2i\Delta_n+\gamma_1)\sigma_{-}+i(\lambda_{k}d_k - g_{cp}c)\sigma_z + \sqrt{2\gamma_1}a_{in}(t),\nonumber\\
\dot d_{k}&=&-(i\Delta_{ex}+\gamma_2)d_k + i\lambda_{k}\sigma_{-}, 
\label{H-eqns}
\end{eqnarray}
where $\gamma_2$, $\Gamma_m$, and $\gamma_1$ denote the decays of the electron-hole excitation relaxation rate, mechanical mode of mirror $M_2$, and G-mode phonon, respectively. 
Furthermore, it should be noted that the system with a zero mean value $\langle\zeta(t)\rangle=0$ is affected by the Langevin noise operator, also known as the Hermitian Brownian noise operator.

Additionally, we examine the input vacuum noises associated with the cavity field $c_{\text{in}}(t)$ along with the atom $a_{\text{in}}(t)$. 
When $\langle c_{\text{in}}(t)\rangle=\langle a_{\text{in}}(t)\rangle=0$, the average values of $c_{\text{in}}(t)$ along with $a_{\text{in}}(t)$ are zero.
Moreover, the atom follows the nonvanishing commutation relations $\langle c_{\text{in}}(t) c_{\text{in}}^{\dagger}(t^{'})=\delta(t-t^{'})\rangle$ as well as $\langle a_{\text{in}}(t) a_{\text{in}}^{\dagger}(t^{'})=\delta(t-t^{'})\rangle$, along with how input vacuum noises influence the cavity field \cite{genes2008emergence}.

The first order in fluctuations is linearized as $O=O_s +\delta O$, where $ O \in (c, b, \sigma$, and $d)$. If we consider nonfluctuation terms in a steady-state condition, the mean values of the observables are given
\begin{eqnarray}
c_{s} &=& \frac{\mathcal{E}_l}{\kappa+i\Delta_2-\frac{g_{cp}^2(i\Delta_{ex}+\gamma_2)}{-(2i\Delta_n+\gamma_1)(i\Delta_{ex}+\gamma_2)+\lambda_{k}^2}},\nonumber\\
b_s &=& \frac{i g_{mc}}{(i\Delta_b + \Gamma_m)} |c_{s}|^2,\nonumber\\
\sigma_s &=& \frac{i\lambda_k}{2i\Delta_n + \gamma_1} d_s - \frac{ig_{cp}}{2i\Delta_n + \gamma_1} c_s,\nonumber\\
d_s &=& \frac{i\lambda_k}{i\Delta_{ex}+\gamma_2} \sigma_s,
\label{steady-state}
\end{eqnarray}
where $\Delta_2=\Delta_c- \frac{g_{mc}^2 |c_s|^2}{m \hbar \omega_m^2}$ is the effective cavity detuning.

Next we introduce the slowly varying amplitudes ($\delta c = \delta c e^{-i\Delta_c t}, \delta b = \delta b e^{-i\omega _b t}, \delta \sigma_- = \delta \sigma_- e^{-2i\Delta_n t},$ and $\delta d_k = \delta d_k e^{-i\Delta_{ex} t}$) on the fluctuation part and obtain

\begin{eqnarray}
\delta \dot c &=& -\kappa \delta c + i g_{mc} (\delta b+ \delta b^{\dagger}) \delta c - i g_{cp}\delta \sigma_{-} +\mathcal{E}_pe^{-i\Delta_p t},\nonumber\\
\delta \dot b &=& - \Gamma_m \delta b + ig_{mc} \delta c^{\dagger}\delta c,\nonumber\\
\delta \dot \sigma_{-} &=& - \gamma_1 \delta \sigma_{-} + i(\lambda_{k}\delta d_k + g_{cp}\delta c) \delta \sigma_z ,\nonumber\\ 
\delta \dot d_{k} &=& - \gamma_2 \delta d_k-i\lambda_{k}\delta \sigma_{-}, 
\label{H1}
\end{eqnarray}
where $\Delta_p=\delta-\omega_b$ is the effective detuning. While deriving the above equation, we use the resolved sideband limit $\omega_b \gg \kappa$, and $\omega_b \approx \Delta_c \approx \Delta_{ex} \approx 2 \Delta_n$. Finally, we apply the ansatz $\delta O = O_{-}e^{-i \Delta_p t}+O_{+}e^{i \Delta_p t}$ and after solving sets of coupled equations, we obtain the amplitude of intracavity field $c_-$ oscillating at the probe frequency $\omega_p$ as
\begin{equation}
    c_- = \frac{(\alpha_3\alpha_4+\lambda_k^2)\alpha_2\mathcal{E}_p}{\alpha_2 \alpha_4 (\alpha_1 \alpha_3 + g_{cp}^2) + \alpha_1 \alpha_2 \lambda_k^2+|c_s|^2 g_{mc}^2 (\alpha_3\alpha_4+\lambda_k^2)},
\end{equation}
where $\alpha_1=\kappa- i\Delta_p$, $\alpha_2=\Gamma_m- i\Delta_p$, $\alpha_3=\gamma_1- i\Delta_p$, and $\alpha_4=\gamma_2- i\Delta_p$. 
The amplitude $c_{-}$ represents the linear response of the system to a weak probe field. To relate $c_{-}$ to the output field, the input-output relation of the cavity may be written as~\cite{agarwal2010electromagnetically,walls_quantum_2025}
\begin{equation}
E_{\text{out}}(t) + \mathcal{E}_p e^{-i\Delta_p t} + \mathcal{E}_l = \sqrt{2\kappa}c,
\end{equation}
where
\begin{equation}
E_{\text{out}}(t) = E_{\text{out}}^0 + E_{\text{out}}^+ \mathcal{E}_p e^{-i\Delta_p t} + E_{\text{out}}^- \mathcal{E}_p e^{i\Delta_p t}.
\end{equation}
After solving these equations, we obtain the relation as $E_{\text{out}}^+ = (\sqrt{2\kappa}c_{-}/\mathcal{E}_p) - 1$ that can be measured by using the homodyne technique~\cite{walls_quantum_2025}. For convenience, we define $E_{\text{out}}^+ + 1  = E_{\text{out}}$. 
The real and imaginary parts (i.e., the quadratures) of the normalized output field encode the absorption and dispersion properties of the probe, respectively. Thus, the output field reflects the system’s linear optical response and can be used to compute the optical susceptibility of the intracavity medium, i.e., \( \chi = E_{\text{out}} \), under the assumption of weak probe field~\cite{khan2020investigation, liu2024generating, waseem_ghsmagno}.

\section{Results and discussion}

In this section, we discuss the impact of coupling $g_{mc}$, $g_{cp}$, and $\lambda_{k}$ on the coherent control of photonic SHE, which allows for precise manipulation of the photonic SHE.
We adopt the following experimentally feasible parameters for the system involving graphene-based nanocavities~\cite{hafeez2019optomechanically, tang2010tunable}: $\Gamma_m/2\pi=140$ Hz, $\omega_b/2\pi=10$ MHz, $\Delta_c/2\pi=\Delta_{ex}/2\pi=\omega_b$, $\Delta_n/2\pi=\omega_b/2$, $\kappa/2\pi=\omega_b/30$, $\gamma_1/2\pi=\gamma_2/2\pi=2 $ kHz, $g_{mc}/2\pi=29$ kHz, $L=1$ $\mu$m, $m=78$ ng, and $P_c=200$ nW. 
The following are the parameters for photonic SHE: $\lambda=1064$ nm, $\epsilon_0=1$, $\epsilon_1=\epsilon_3=2.22$, $d_1=d_3=0.1$ $\mu$m, and $d_2=1$ $\mu$m.

\begin{figure}[t]
	\centering
	\includegraphics[width=0.9\linewidth]{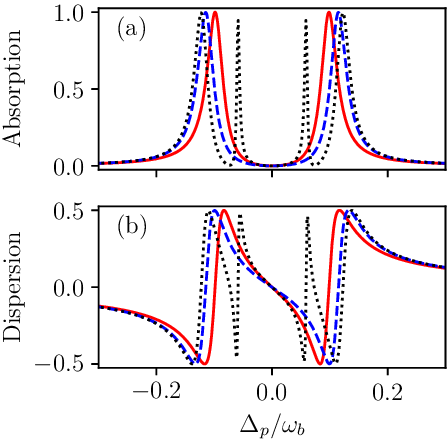}
\caption{ (a) The absorption and (b) dispersion characteristics curves as a function of effective detuning $\Delta_p$. We choose the power of the pump field $P_p=200$ nW. Red curve represents the results at $g_{cp}=0$ and $\lambda_k=0$, for dashed blue curve $g_{cp}=20g_{mc}$ and $\lambda_k=0$, and for dotted black curve $g_{cp}=20g_{mc}$ and $\lambda_k=25g_{mc}$. The rest of the parameters are given in the text.} 
\label{fig2}
\end{figure}

We start our analysis by examining the absorption and dispersion of the probe field as a function of the effective detuning $\Delta_p$, as shown in Fig. \ref{fig2}(a) and (b). The solid red curve ($g_{cp}=\lambda_k=0$) shows the standard optomechanical-induced transparency with zero absorption and dispersion at zero detuning $\Delta_p = 0$. This transparency window appears due to the coupling $g_{mc}$ and the power of the pump field.
Upon introducing coupling ($g_{cp}=20g_{mc}$, $\lambda_k=0$), as shown by the dashed blue curve, the transparency window in the absorption profile becomes broadened. Similarly, the slope of the dispersion profile becomes less steep.
This change happens because the coupling modifies the energy levels of the system.
When both the G-mode phonon and electronic state coupling are introduced ($g_{cp}=20g_{mc}$, $\lambda_k=25g_{mc}$), three transparency windows appear as seen in the dotted black curve in Fig.\ref{fig2}(a). 
These changes highlight the sensitivity of the system's absorption properties to the coupling strengths $g_{cp}$ and $\lambda_k$, which are absent in the previous monolayer case~\cite{akhter}.
We choose the condition $\omega_b=2 \Delta_n$ to derive the output field quadrature. In an unresolved sideband condition, such as $\omega_b=\Delta_n$ will result the Fano-resonances profile in the quadrature of the output field and discussed somewhere else \cite{hafeez2019optomechanically}.

\begin{figure}[t]
\centering
\includegraphics[width=1\linewidth]{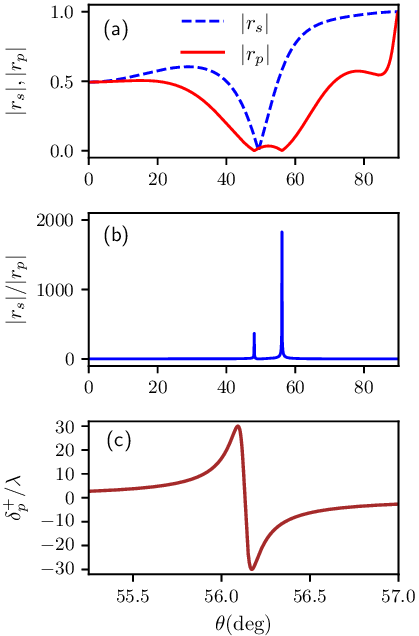}
\caption{(a) Reflection coefficients $|r_s|$ and $|r_p|$, (b) ratio of reflection coefficients $|r_s|/|r_p|$, and (c) photonic SHE as a function of probe field incident angle $\theta$ when $\Delta_p=0$. Here $g_{cp}=\lambda_k = 0$.}
\label{fig3}
\end{figure}

As discussed above, coupling strengths influence the output field's quadrature and permittivity $\epsilon_2$. As a result, reflection coefficients and photonic SHE are modified. 
Therefore, we next analyze $|r_s|$ and $|r_p|$ as a function of the incident angle $\theta$ whose results are shown in Fig. \ref{fig3}(a). For these results, we choose $\Delta_p=0$ and $g_{cp}=\lambda_k=0$. 
\begin{figure}[ht!]
	\centering
	\includegraphics[width=1\linewidth]{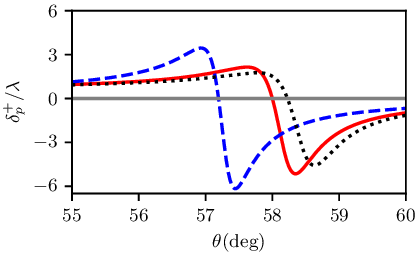}\\
\caption{Photonic SHE as a function of $\theta$ at probe field detuning $\Delta_p=0.03 \omega_b$. Red curve represents the results at $g_{cp}=0$ and $\lambda_k=0$, dashed blue curve  shows results at $g_{cp}=20g_{mc}$ and $\lambda_k=0$, and dotted black curve shows result at $g_{cp}=20g_{mc}$ and $\lambda_k=25g_{mc}$.}
\label{fig4}
\end{figure}
At approximately $\theta \approx 50^\circ$, both $|r_s|$ and $|r_p|$ are zero. As the angle increases, $|r_p|$ remains minimal, while $|r_s|$ starts to increase significantly around $\theta \approx 56.2^\circ$, which is called the Brewster angle.
As a result, in Fig. \ref{fig3}(b), their ratio $|r_s|/|r_p|$ significantly enhances at $\theta \approx 56.2^\circ$ around 2000, which we call resonance condition. As is evident from Eq.(~\ref{halleffect}), a larger ratio of $|r_s|/|r_p|$ can lead to enhanced photonic SHE around $\theta \approx 56.2^\circ$, which is shown in Fig.~\ref{fig3}(c). The photonic SHE $\delta^+_p$ changes sign from a maximum positive value of $30 \lambda$ to a maximum negative value $-30 \lambda$ around the incident angle $\theta \approx 56.2^\circ$ (Brewster angle). This change of sign appears to the $\pi$ phase alteration of the relative phase between $|r_s|$ and $|r_p|$~\cite{waseem2024gain}.
It may be noted that $\delta^+_p=\pm 30 \lambda$ is half of the incident Gaussian beam waist of $w_0=60 \lambda$. 
The results of photonic SHE are similar when $g_{cp} \neq 0$ and $\lambda_k \neq 0$ due to zero absorption and dispersion at $\Delta_p=0$, and are not shown for simplicity.

Next, we examine the photonic SHE $\delta_p^+$ of the reflected probe beam as a function of $\theta$ in Fig. \ref{fig4} at slightly nonzero detuning $\Delta_p =0.03 \omega_b$. 
The solid red curve represents the results at $g_{cp}=0$ and $\lambda_k=0$.
The solid curve shows that the peak values of $\delta_p^+$ appear at two degrees higher ($\theta \approx 58.2^\circ $) compared with the $\Delta_p=0$ case.
The dashed blue curve shows the result for $g_{cp}=20g_{mc}$ and $\lambda_k=0$. 
The magnitude of photonic SHE increases and appears at $\theta \approx 57.2^{\circ}$ because of the broadening in the transparency window.
\begin{figure}[ht!]
    \centering
    \includegraphics[width=1.0\linewidth]{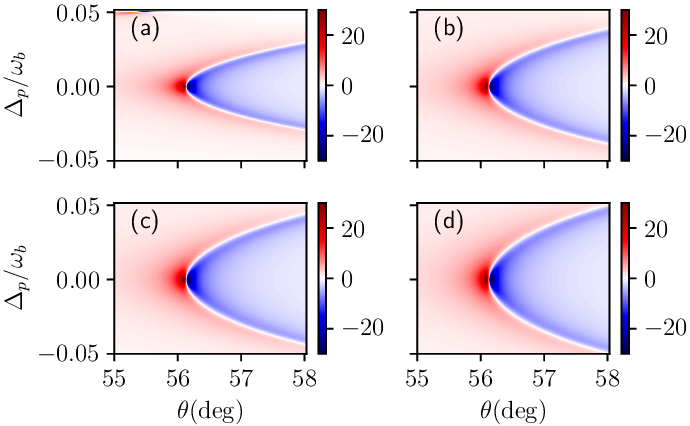}    
    \caption{Photonic SHE ($\delta^+_p/\lambda$) as a function incident angle of the probe field $\theta$ and probe field effective detuning $\Delta_p$ at $\lambda_k=0$. For (a) $g_{cp}=0$, (b) $g_{cp}=20g_{mc}$, (c) $g_{cp}=25g_{mc}$, and (d) $g_{cp}=30g_{mc}$. The rest of the parameters are unchanged.}
    \label{fig5}
\end{figure}
The dotted black curve shows the result for $g_{cp}=20g_{mc}$ and $\lambda_k=25g_{mc}$. Since the coupling $\lambda_k$ shrinks the transparency window near zero probe field detuning and increases the absorption in the nearby region. As a result, the magnitude of the photonic SHE decreases and appears at an angle slightly higher than  $\theta \approx 58.2^{\circ} $.
In all three results, one common feature is the asymmetry in the photonic SHE profile. In other words, the peak positive value is smaller than the peak negative value. 
As $\Delta_p$ increases, the resonance conditions shift, causing the peak of $\delta_p^+$ to change from incident resonance angles $\theta \approx 56.2^{\circ }$. When $g_{cp} \neq 0$ and $\lambda_k \neq 0$, photonic SHE is nonidentical, unlike the case of $\Delta_p=0$.    
This shift indicates a tunable control of photonic SHE by adjusting the coupling terms, which can be used to achieve ultrasensitive detection of refractive index variations or local field perturbations.
%
\begin{figure*}[t!]
    \centering
    \includegraphics[width=1\linewidth]{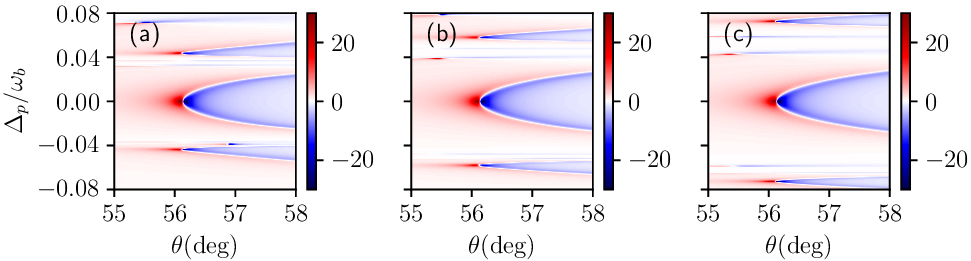}
    \caption{Photonic SHE ($\delta^+_p/\lambda$) as a function incident angle of the probe field $\theta$ and probe field effective detuning $\Delta_p$. In (a) $\lambda_k=15g_{mc}$, (b) $\lambda_k=20g_{mc}$, and (c) $\lambda_k=25g_{mc}$. We select $g_{cp}=20g_{mc}$ and same other parameters.}
    \label{fig6}
\end{figure*}

Now, we explore the effects of cavity-phonon coupling strength $g_{cp}$ at $\lambda_k=0$. 
Figure~\ref{fig5} shows density plots of $\delta_p^+$ as a function of both $\theta$ and $\Delta_p/\omega_{b}$ for Figs.~\ref{fig5}(a) $g_{cp}=0$,~\ref{fig5}(b) $g_{cp}=20g_{mc}$,~\ref{fig5}(c) $g_{cp}=25g_{mc}$, and~\ref{fig5}(d) $g_{cp}=30g_{mc}$.
In all cases, $\delta_p^+$ peaks positive $30 \lambda$ and change sign to peak negative $-30 \lambda$ around $\theta \approx 56.2^\circ$ at $\Delta_p \approx 0$.
The white parabolic curves show the Brewster angle position around which photonic SHE switches the sign.
When $\Delta_p \neq 0$, the photonic SHE from positive to negative becomes asymmetrical, similar to what is observed in Fig.~\ref{fig4}. The magnitude of the negative photonic SHE appears larger than the positive photonic SHE at a higher incident angle. Higher negative photonic SHE appears as the detuning range increases on either side of $\Delta_p = 0$. 
In other words, photonic SHE is mostly negative at larger nonzero detuning and incident angles higher than $\theta \approx 56.2^{\circ}$.
The range of detuning at which negative photonic SHE appears gets larger due to broadening in the transparency window when the magnitude of coupling $g_{cp}$ increases.
Looking at fixed nonzero detuning, it is evident that the maximum positive and maximum negative magnitude of photonic SHE increase with the increase of $g_{cp}$. This change of magnitude is more pronounced at smaller detuning. Similarly, the angular position of the Brewster angle (shown by white line curves in the density plot), maximum positive $\delta_p^+$, and maximum negative $\delta_p^+$ change to a value higher than $\theta \approx 56.2^{\circ}$. For example, at $\Delta_p=0.03 \omega_b$, the angular position of the maximum negative $\delta_p^+$ appears at $\theta \approx 58.4^{\circ}$, which decreases with the increase of $g_{cp}$. This means a change in magnitude and angular position of photonic SHE can be used for sensing the cavity field and G-mode phonon coupling strength.
In short, as the coupling strength $g_{cp}$ increases, the sensitivity of the system to photonic SHE also increases with respect to both the incident angle $\theta$ and the detuning $\Delta_p$.
The broadening in photonic SHE occurs due to increased energy exchange between the probe field and the bilayer graphene optomechanical system, leading to a wider angular response. Moreover, the negative shift in photonic SHE at nonzero detuning can be attributed to asymmetric energy redistribution, where detuning alters the balance of energy transfer, shifting the polarization response.

Finally, we include the effect of $\lambda_k$ together with $g_{mc}$ and $g_{cp}$ on photonic SHE. We plot the results of photonic SHE as a function of $\theta$ and $\Delta_p/\omega_b$, which are shown in Fig.~\ref{fig6}.
In Figs.~\ref{fig6}(a) $\lambda_k=15g_{mc}$,~\ref{fig6}(b) $\lambda_k=20g_{mc}$,~\ref{fig6}(c) $\lambda_k=25g_{mc}$ and the fixed parameter $g_{cp}=20g_{mc}$.
In all cases, $\delta_p^+$ peaks positive and changes the sign to peak negative at $\theta \approx 56.2^{\circ}$ at three different values of detuning $\Delta_p$ because of three transparency windows in the absorption profile. 
Here, we would like to mention that the sidebands in the photonic SHE response originate from the coupling of the G-mode phonon with electron-hole excitation. 
The peak value of photonic SHE at sidebands is slightly lower than the peak value at $\Delta_p=0$.
The range of detuning at which the photonic SHE changes sign from positive to negative is narrower for side bands compared with $\Delta_p=0$. 
The sidebands of photonic SHE appear farther away from the central band as the value of $\lambda_k$ increases.
When $\lambda_k \neq 0$ and $g_{cp}=0$, we obtain results exactly similar to the results obtained in Fig.~\ref{fig5}(a).
In other words, photonic SHE is independent of $\lambda_k$ when $g_{cp}=0$.  This happens because electron-hole excitations are not directly coupled to the nanocavity field but rather indirectly via G-mode phonon interactions.
The direct coupling between the cavity field and electron-hole excitations is negligible, as the cavity photon energy ($\sim$ 1.17 eV at 1064 nm) is far detuned from the maximum tunable bandgap in dual-gated bilayer graphene (0.2 to 0.25 eV)\cite{zhang_direct_2009}. This significant energy difference justifies neglecting such interactions in our model.
Furthermore, including direct cavity–electron-hole coupling would require a different theoretical approach and is beyond the scope of the current work.

\section{Conclusion}
In conclusion, we theoretically explored the photonic SHE as a function of probe field detuning and incident angle in dual-gated bilayer graphene inside an optomechanical cavity.
Photonic SHE can be tuned using cavity optomechanical coupling $g_{mc}$, cavity field and G-mode phonon interaction $g_{cp}$, and coupling between G-mode phonon and electronic states $\lambda_{k}$.
Our results are equally valid for a standard cavity optomechanical system~\cite{agarwal2010electromagnetically} in the absence of cavity field and G-mode phonon interaction. 
Notably, the photonic SHE is most pronounced near the critical angle of approximately $56^{\circ}$ at probe field detuning $\Delta_{p}=0$ for all coupling strengths. 
Photonic SHE changes to asymmetric for each coupling strength at $\Delta_{p} \neq 0$ and incident angle higher than $56^{\circ}$. 
The result obtained also indicates that coupling strengths can be estimated using photonic SHE.
The dual-gated configuration of bilayer graphene inside an optomechanical cavity may provide an effective platform for realizing tunable photonic SHE, making it a promising candidate for next-generation hybrid photonic sensors.\\

In this study, we consider linear coupling between the cavity field and the G-mode phonon in bilayer graphene. A perpendicular electric field breaks the inversion symmetry between the upper and lower graphene layers, inducing different charge distributions on the carbon atoms in each layer. As a result, these atoms behave like oppositely charged ions in a polar crystal, and their phonon vibrations become predominantly infrared-active~\cite{tang2010tunable,gatevolt}. 
Dual-gated bilayer graphene routinely reaches displacement fields $0.3-1.0~\rm{V}/\rm{nm}$~\cite{gatevolt}. 
In this range, the infrared G-phonon coupling is very strong~\cite{gatevolt,zhang_direct_2009}, but Raman activity generally persists~\cite{ferrari_raman_2013}.
Furthermore, field-dependent screening, doping~\cite{cappelluti_charged-phonon_2012}, gating effects~\cite{malard_observation_2008}, and optomechanical effects can modulate this Raman activity rather than eliminate it.
Recent studies investigated the hybridization between an optical cavity mode and a Raman-active phonon mode in bilayer graphene and other two-dimensional materials~\cite{ojeda_collado_equilibrium_2024, bourzutschky_raman-phonon-polariton_2024}.
Such Raman-active coupling may occur resonantly via excitons~\cite{bourzutschky_raman-phonon-polariton_2024} or off-resonantly through second-order processes involving parametric amplification~\cite{ojeda_collado_equilibrium_2024}. 
This study serves as an initial step toward extending Raman-active cavity-phonon coupling in the optomechanical systems and photonic SHE.

We would like to mention that our proposed theoretical model is motivated by previous experiments~\cite{tang2010tunable, yan_tunable_2014,weis_optomechanically_2010}. 
Interestingly, phonon-induced transparency and Fano resonances have been reported in bilayer graphene~\cite{tang2010tunable, yan_tunable_2014}, while optomechanically induced transparency was demonstrated in Ref.~\cite{weis_optomechanically_2010}.
Motivated by these experiments in two different areas, we consider bilayer graphene inside an optomechanical cavity and investigate the photonic SHE.
We believe that the results of our theoretical study on bilayer graphene in an optomechanical setup may serve as a basis for future experimental investigations using quantum weak measurements~\cite{lee_real-time_2025}.

\section*{Acknowledgements}\label{section:Acknowledgements}
This work was supported by the National Natural Science Foundation of China (Grant No. 12174301), the Natural Science Basic Research Program of Shaanxi (Program No. 2023-JC-JQ-01), and the Fundamental Research Funds for the Central Universities. Muhammad Waseem acknowledged the fruitful discussion with Dr. Muhammad Irfan and Dr. Muzamil Shah during INSC 2024.

\bibliographystyle{apsrev4-2}
\bibliography{Ref}
\end{document}